\documentclass[11pt]{amsart}

\usepackage{amsmath}
\usepackage{palatino}

\usepackage{ tikz, pgfplots, todonotes,ulem,mathrsfs}
\definecolor{darkblue}{rgb}{0.1,0.1,0.45}
\usepackage{caption}
\usepackage{amssymb}
\usepackage[colorlinks=true, pdfstartview=FitV, linkcolor=darkblue, citecolor=black, urlcolor=blue]{hyperref}
\definecolor{shadecolor}{rgb}{0.9, 0.9, 0.81}
\def\tt{\tilde{t}}

\def\a{\alpha}
\def\g{\gamma}

\def\b{\beta}
\def \res#1{ \mathop{\rm res}_{#1}}
\def \le{\left}
\def\ri{\right}

\def\d{\delta}

\def\l{\lambda}

\def \Id { \, {\rm Id}}

\def\O{\Omega}

\def\kk{q}

\def\CP1{{\mathbb C}P^1}
\def\Pcal{{\cal P}}

\def\Rcal{{\mathcal R}}

\def\C{{\mathbb C}}

\def\la{\label}

\def\c{\cite}
\def\f{\frac}

\def\p{\partial}

\def\det{{\rm det}}
\def\gh{{\hat{g}}}
\def\0{S}
\def\log{\ln}

\def\la{\label}

\def\c{\cite}
\def\f{\frac}

\def\p{\partial}

\def\0{S}
\def\1{T}

\def\log{\ln}

\def\det{{\rm det}}

\def\xh{\hat{x}}
\def\Mcal {{\mathcal M}}

\def \eqref #1{ (\ref{#1})}

\newlength{\dinwidth}
\newlength{\dinmargin}
\def \&{\hspace{-18pt}&}

\def\QED{ {\hfill $\blacksquare$}\par \vskip 4pt}
\usepackage{framed,color}

\def\Pcal{\mathcal P}
\def\lt{\tilde{l}}
\def\ct{\tilde{c}}

\newtheorem{theorem}{Theorem}[section]

\newtheorem{proposition}[theorem]{Proposition}

\newtheorem{lemma}[theorem]{Lemma}

\def\nn{\nonumber}

\def\be{\begin{equation}}
\def\ee{\end{equation}}
\def\ben{\begin{displaymath}}
\def\een{\end{displaymath}}
\def\baa{\begin{eqnarray}}
\def\eaa{\end{eqnarray}}

\def\ba{\begin{array}}
\def\ea{\end{array}}
\def\la{\label}
\def\p{\partial}
\def\Pcal{{\mathcal P}}
\def\Hcal{{\mathcal H}}

\def\zetah{\hat{\zeta}}

\def\xh{\hat{\xi}}

\def\Ch{\hat{C}}

\makeatletter
\@addtoreset{equation}{section}
\makeatother

\def\C{{\mathbb C}}
\def\R{{\mathbb R}}

\def\O{\Omega}
\def\Oh{{\widehat{\Omega}}}

\def\Qcal{{\mathcal Q}}
\def\Mcal{{\mathcal M}}

\def\Mcal{{\mathcal M}}

\def\f{\frac}

\def\d{\delta}
\def\a{\alpha}
\def\b{\beta}
\def\g{\gamma}

\def\CC{C}

\def\Ch{\widehat{C}}

\def\gh{\hat{g}}

\def\Acal{{\mathcal A}}

\def\l{\lambda}

\def\Oh{\hat{\Omega}}
\def\gh{\hat{g}}

\def\dim{{\rm dim}}

\def\nn{\nonumber}

\def\be{\begin{equation}}
\def\ee{\end{equation}}
\def\ben{\begin{displaymath}}
\def\een{\end{displaymath}}
\def\baa{\begin{eqnarray}}
\def\eaa{\end{eqnarray}}

\def\ba{\begin{array}}
\def\ea{\end{array}}
\def\la{\label}
\def\p{\partial}

\makeatletter
\@addtoreset{equation}{section}
\makeatother

\def\C{{\mathbb C}}
\def\R{{\mathbb R}}

\def\2x2{{\left(\!\!\begin{array}{cc}a&b\\c&d\\\end{array}\!\!\right)}}

\def\f{\frac}

\def\d{\delta}
\def\a{\alpha}

\def\p{\partial}

\def\CC{C}
\def\Ch{{\hat{C}}}

\def\f{\frac}
\def\l{\lambda}

\def\Acal{{\mathcal A}}
\def\a{\alpha}
\def\b{\beta}
\def\p{\partial}

\def\Ch{\widehat{C}}

\def\la{\label}

\usepackage{geometry}                
\usepackage{graphicx}
\usepackage{amssymb}
\usepackage{epstopdf}

\title{Spaces of Abelian differentials  and  Hitchin's spectral covers}

\begin{document}
\maketitle
\vspace{0.2cm}
\begin{center}
\begin{Large}
\fontfamily{cmss}
\fontsize{17pt}{27pt}
\selectfont
\textbf{}
\end{Large}\\
\bigskip
M. Bertola$^{\dagger\ddagger}$\footnote{Marco.Bertola@\{concordia.ca, sissa.it\}},  
D. Korotkin$^{\dagger}$ \footnote{Dmitry.Korotkin@concordia.ca},
\\
\bigskip
\begin{small}
$^{\dagger}$ {\it   Department of Mathematics and
Statistics, Concordia University\\ 1455 de Maisonneuve W., Montr\'eal, Qu\'ebec,
Canada H3G 1M8} \\
\smallskip
$^{\ddagger}$ {\it  SISSA/ISAS,  Area of Mathematics\\ via Bonomea 265, 34136 Trieste, Italy }\\
\end{small}
\vspace{0.5cm}
{\bf Abstract} \end{center}

Using the embedding of the moduli space of generalized $GL(n)$ Hitchin's spectral covers to the moduli space of 
meromorphic Abelian differentials we  study the variational formul\ae\  of the period matrix, the canonical bidifferential, the prime form 
and the Bergman tau function. This leads to residue formul\ae\  which  generalize the Donagi-Markman formula for variations of the period matrix. The computation of second derivatives of the period matrix reproduces the formula derived in \cite{Baraglia} using the framework of topological recursion.

\tableofcontents

\section{Introduction}

The geometry of spaces of Abelian differentials on Riemann surfaces has attracted interest in relationship with the theory of Teichm\"uller flow
\cite{KZ,KZ2,EKZ}. Methods inspired by the theory of integrable systems were applied to the study of these spaces in \cite{JDG,MRL,CMP} where an appropriate version of deformation theory of Riemann surfaces and the formalism of tau functions  was developed. In particular, variations of moduli and of various canonical objects associated to a Riemann surface were computed in \cite{JDG} (holomorphic case) and in \cite{CMP} (meromorphic case). The
{\it Bergman tau function} introduced in \cite{JDG} is a natural generalization of Dedekind's eta-function to higher genus.

The origin of Hitchin's spectral covers and their moduli spaces is the dimensional reduction of self-dual Yang-Mills equations on a  four-dimensional space represented as the  product of a Riemann surface and $\R^2$ \cite{Hitchin1}. Such  a  dimensional reduction gives a family of completely integrable systems
associated to families of Riemann surfaces of arbitrary genus \cite{Hitchin2}. Hamiltonians of such integrable systems (we consider here only the $GL(n)$ gauge group) are encoded in the $n$-sheeted {\it spectral cover} of a Riemann surface.  The moduli space of spectral covers for a base Riemann surface of given genus  was also intensively studied (see \cite{Atiyah,DonagiMarkman}). In particular, the {\it Donagi-Markman cubic}
describes variations of the period matrix of the spectral cover for fixed base, answering the question posed in \cite{Atiyah}. Variations of the canonical meromorphic bi-differential on these spaces were derived in \cite{Baraglia} using the formalism developed in \cite{Fay92}.

The space of Hitchin's spectral covers admits a natural embedding in a space of Abelian differentials; this embedding was used in \cite{Faddeev} 
to define a natural version of Bergman tau functions on spaces of spectral covers (with variable or fixed base) and find the class of the locus of degenerate covers (the universal Hitchin's
discriminant) in the Picard group of the universal moduli space of spectral covers.

In this paper we further exploit this embedding to show how variational formul\ae\  for the  period matrix, the canonical bidifferential and the prime form on the moduli spaces
of generalized Hitchin's systems (when the coefficients of the equation defining the spectral cover are allowed to be meromorphic differentials)
can be deduced from variational formul\ae\  on moduli spaces of meromorphic Abelian differentials derived in \cite{JDG,CMP}. In the special case of 
regular Hitchin's systems we reproduce residue formul\ae\  for the canonical bidifferential obtained in \cite{Baraglia} and for the period matrix 
(given by the Donagi-Markman cubic \cite{DonagiMarkman}).  We also derive residue formul\ae\  for variations of Bergman tau function of spaces of spectral covers 
for the holomorphic case.

The formulas for the  second derivatives of the period matrix (in holomorphic case) found in our formalism coincide with expressions derived in \cite{Baraglia} using the formalism of topological recursion of \cite{EO}. These formul\ae\  are rather cumbersome in contrast to analogous formul\ae\  on spaces of Abelian differentials. This suggest a possibility of existence of a natural simple structure on spaces of Abelian differentials which underlie the topological recursion framework on spaces of spectral covers.


\section{Spaces of generalized spectral covers}
\la{sgs}

Denote by $\CC$ a Riemann surface of genus $g$, with  $m$ marked points $y_1,\dots,y_m$ on $\CC$ and 
associated corresponding
multiplicities $k_1,\dots,k_m$, $k_j\geq 1$. 
The Higgs bundle on $\CC$ is a pair $(E,\Phi)$ where $E$ is a holomorphic vector bundle and $\Phi$ (the {\it Higgs field}) is a holomorphic (or meromorphic, depending on the specific setting) $Ad_E$-valued $1$-form on $\CC$ \cite{Hitchin2,DonagiMarkman}. For a given base curve $\CC$ and a degree of the bundle $E$ the space of pairs $(E,\Phi)$ is called the moduli space of Higgs bundles.

Consider a meromorphic $GL(n)$ Higgs field $\Phi$ with  poles at  $y_j$'s of the corresponding order  $k_j$,  $j=1,\dots,m$. 
We also assume a generic form of the singular parts of $\Phi$ near these poles. 
The {\it spectral curve} $\Ch$ is defined {as a locus in $T^* C$} by the equation $\det(\Phi-v\Id)=0$, which  can be written as   
\be
v^n+Q_{1}v^{n-1}+\dots+Q_n=0
\la{spcov}
\ee
where $Q_\ell$ is a meromorphic $\ell$-differential on $\CC$ with pole of order $\ell  k_j$ at the point $y_j$ thanks to the genericity assumption.

For fixed $\CC$ and $\{y_j\}_{j=1}^m$ we denote by $\Mcal_H^n[{\bf k}]$
the moduli space of  curves (\ref{spcov}) which can be identified with the moduli space of sets of the differentials $Q_\ell$ with poles of appropriate order at the points $y_j$. Namely, denoting by $\Omega_\ell$ the vector space of 
$\ell$-differentials on $\CC$ with poles of order $\ell k_j$ at $y_j$, we have
$$
\Mcal_H^n[{\bf k}]=\bigoplus_{\ell=1}^n \Omega_\ell
$$

Denote by $\pi$ the projection $\Ch\to\CC$. Assuming that the branch points of $\Ch$ do not coincide with $y_j$ 
we have $\pi^{-1}(y_j)=\{y_j^{(s)}\}_{s=1}^{n}$. 

The meromorphic Abelian differential $v$ has, on $\Ch$, poles of order $k_j$ at all $y_j^{(s)}$.
Denote by $\chi_j$ a local coordinate on $\CC$ near $y_j$; since we have assumed  that $y_j$ is a not a branch point of $\Ch$ we can use $\chi_j$ also as local coordinate near each $y_j^{(s)}$ for $s=1,\dots,n$. Consider the singular parts of $v$ at $y_j^{(s)}$:
\be
v(\zeta_j)= \left(\f{C_j^{(s),k_j}}{\chi_j^{k_j}}+\f{C_j^{(s),k_j-1}}{\chi_j^{k_j-1}}+\dots+\f{C_j^{(s),1}}{\chi_j}+ O(1)\right)d\chi_j\;.
\la{singv}
\ee

The discriminant $W$ of the equation (\ref{spcov})  is a meromorphic $n(n-1)$ differential on $\CC$ which has 
pole of order $n(n-1)k_j$ at $y_j$. Therefore, the total degree of poles of $W$ is $n(n-1)\sum_{j=1}^m k_j$ and the number of its zeros (i.e. the number of branch points of $\Ch$) is 
\be
p=n(n-1)\le(2g-2+ \sum_{j=1}^m k_j\ri)\;.
\la{nubp}\ee
It follows from the Riemann--Hurwitz formula that  the genus of $\Ch$ equals
\be
\gh=n^2(g-1)+1+\f{n(n-1)}{2}\sum_{j=1}^m k_j
\ee
The degree of the divisor of zeroes of the Abelian differential $v$  on $\Ch$ is
\be
r=2\gh-2+n\sum_{j=1}^m k_j
\la{defr}
\ee
The dimension of  $\Mcal_H^n[{\bf k}]$ equals to the sum of dimensions of spaces of coefficients of (\ref{spcov}), which is computed as 
$$
\left(\sum k_j -1+g\right)+\left(2\sum k_j +3(g-1)\right)+\dots+\left(n\sum k_j +(2n-1)(g-1)\right).
$$
Assuming that at least one $k_j>0$, the above gives 
\be
\dim \Mcal_H^n[{\bf k}]=\f{n(n+1)}{2}\sum k_j +n^2(g-1)=\gh+n\sum_{j=1}^m k_j -1\;.
\la{dimM}
\ee
On  the moduli space $\Mcal_H^n[{\bf k}]$ we introduce the following local coordinates:
\be
\left\{ \{A_\a\}_{\a=1}^{\gh}, \{C_j^{(s),\ell }\}, j=1,\dots,m, \;s=1,\dots,n,\; \ell =1,\dots,k_j,\; (j,s,\ell )\neq (1,1,1)\right\}
\la{coord}
\ee
where $C_j^{(s),\ell }$ are coefficients in singular parts of $v$ near $y_j^{(k)}$ (\ref{singv}) (these coefficients of course depend on the choice of local coordinates $\chi_j$ near $y_j$ on $\CC$), and $A_\a$ are $a$-periods of $v$
under an arbitrary choice of Torelli marking:
\be
A_\a=\int_{a_\a}v\;.
\ee
The coefficient $C_1^{(1),1}$ is not an independent coordinate since the sum of residues of $v$ on $\Ch$ vanishes:
\be
\sum_{j=1}^m\sum_{s=1}^{k_j}C_j^{(s),1}=0\;.
\ee
We observe that the number of coordinates (\ref{coord}) coincides with the dimension  (\ref{dimM})  of   $\Mcal_H^n[{\bf k}]$.

Subordinate to the choice of  Torelli marking we also define the normalized first-kind Abelian differentials (holomorphic) $v_\a$ with the  property  
\be
\oint_{a_\beta} v_\a = \delta_{\alpha \beta}.
\label{1stkind}
\ee
We similarly  define the normalized second-kind differentials $w_j^{(s),l}$ on $\Ch$  with prescribed singular part:
\be
\oint_{a_\alpha} w_j^{(s),\ell} = 0,\qquad
w_j^{(s),\ell }(x)=\left(\f{1}{\chi_j^{l}}+O(1)\right) d\chi_j\;,\hskip0.7cm x\sim y_j^{(s)}\;,\hskip0.7cm \ell =2,\dots, k_j
\la{second}
\ee
and the normalized differentials of the third kind $u_j^{(s)}(x)$ on $\Ch$ which have  simple poles at $y_1^{(1)}$ and $y_j^{(s)}$ with residues $-1, +1$, respectively.

Since the moduli of the base curve $\CC$ are kept constant, we can define unambiguously the derivative with respect to the moduli of our space for  any Abelian differential $w$ on $\Ch$. To wit, we fix a local chart $D$ on $\CC$ with a local coordinate $\xi$ and lift $D$
to all sheets of $\Ch$. Then in any connected component of $\pi^{-1}(D)$ we can use $\xi$ as a local coordinate away from ramification points. We express the differential $w$  in such  coordinate $w=f(\xi) d\xi$ and define
\be
\f{d w}{d z_k}=\f{df(\xi)}{dz_k} d\xi
\la{defder}\ee
where the coordinate $\xi$ remains fixed under differentiation. Clearly, the definition (\ref{defder}) is independent  of the choice of the local coordinate $\xi$ because the moduli of the base curve are kept constant. Keeping this in mind we formulate the following proposition.

\begin{proposition}\la{propv}
The following variational formul\ae\  of $v$ with respect to coordinates (\ref{coord}) on $\Mcal_H^n[{\bf k}]$ hold:
\be
\f{\p v}{\p A_\a}= v_\a\;,
\la{vA}
\ee
\be
\f{\p v}{\p C_j^{(s),l}}= w_j^{(s),\ell } \;,\hskip0.8cm    \ell =2,\dots, k_j
\la{vC1}
\ee
where $w_j^{(s),\ell }$ are normalized ( i.e. with $\int_{a_\alpha} w_j^{(s),\ell}=0$) differentials of second kind defined by (\ref{second})
and
\be
\f{\p v}{\p C_j^{(s),1}}= u_j^{(s)}
\la{vC2}
\ee
where $j=1,\dots,m$ and $s=1,\dots,n$; $u_j^{(s)}(x)$ are the normalized differentials of the third kind on $\Ch$ defined after (\ref{second}).
\end{proposition}

{\it Proof.} First notice that the differential $v$ vanishes at all branch points $x_j$ of $\Ch$; generically these zeros are of first
order. This is due the fact that a coefficient $Q_k$ of equation (\ref{spcov}) is a $k$-differential on $\CC$. Being lifted from $\CC$ to $\Ch$, it gains a zero of order $k$ at each branch point since near the ramification point $x_j$ the local coordinate on $\Ch$ is given by $(\xi-\xi_j)^{1/2}$
where $\xi$ is the local coordinate on $\CC$ near $\pi(x_j)$ ($\xi$ is assumed to be independent of coordinates (\ref{coord})) and $\xi_j=\xi(\pi(x_j))$. In particular, the $n$-differential $Q_n$, being lifted to $\Ch$, has zeros of order $n$ at all branch points (as well as zeros lifted to $\Ch$ from its zeros on $\CC$).

Therefore, locally near $x_j$,t we have
$$
v(\xi)=(\xi-\xi_j)^{1/2}(a_0+a_1 (\xi-\xi_j)^{1/2}+\dots) d(\xi-\xi_j)^{1/2}=\f{1}{2}(a_0+a_1 (\xi-\xi_j)^{1/2}+\dots)d\xi\;.
$$
Although $\xi$ is independent of the  moduli coordinates (\ref{coord}), the coordinate $\xi_j$ of the branch point $\pi(x_j)$ does depend on them, and differentiation with respect to any coordinate $z$ from the list (\ref{coord}) gives
$$
\f{\p v}{\p z} =\f{1}{4}\left(-\f{a_1(\xi_j)_z}{(\xi-\xi_j)^{1/2}}+ O(1)\right)d\xi=-\f{1}{2}(a_1(\xi_j)_z+o(1))d\sqrt{\xi-\xi_j}
$$
which is holomorphic (although generically non-vanishing) at $x_j$.
It then follows that all the  differentials ${\p v}/{\p z}$ are holomorphic at the branch points, and can have poles only at the  $y_j^{(s)}$'s. 

The differentials $\p v/\p A_j$ are holomorphic since the coefficients of the singular parts of $v$ near all $y_i^{(s)}$
are independent of $A_j$. Moreover, all $a$-periods of $\p v/\p A_j$ vanish except for the period over $a_j$, which equals $1$.
Therefore, we deduce  (\ref{vA}).

Consider $\p v/\p C_j^{(s),\ell }$ for $l\geq 2$. The only singularity of this differential is at   $y_j^{(s)}$ and its singular part there 
coincides with the one of $w_j^{(s)}$. Moreover, since the  $A_\alpha$  and the $C_j^{(s),\ell }$ coordinates  are independent of each other,
all $a$-periods of $\p v/\p C_j^{(s),\ell }$ vanish; thus this differential coincides with $w_j^{(s)}$.

Similarly, one verifies that the differential  $\p v/\p C_j^{(s),1}$ coincides with the third kind differential $u_j^{(s)}$.
\QED

We are going to combine this proposition with the variational formul\ae\  on moduli spaces of meromorphic Abelian differentials obtained in \cite{JDG,CMP} which we discuss next.

\section{Variational formul\ae\  and Bergman tau function on moduli spaces of meromorphic Abelian differentials}


Denote by $\Hcal_{\gh}[d_1,\dots,d_\kk]$ the moduli space of pairs $(\Ch,v)$ where $\Ch$ is a Riemann surface
of genus $\gh$ and $v$ is a meromorphic differential on $\Ch$ with $\kk$ poles $y_1,\dots,y_\kk$ of orders $d_1,\dots,d_\kk$,
respectively, and simple zeros $x_1,\dots,x_r$ where $r=2\gh-2+\sum_{i=1}^\kk d_i$. The notations $\Ch$ and $\gh$ are now used  in agreement with the previous discussion.
The dimension of $\Hcal_{\gh}[d_1,\dots,d_\kk]$ is the sum of: $3\gh-3$ moduli parameters of $\Ch$, $\kk$  positions of the singularities, $\sum_{j=1}^\kk d_j -1$ coefficients of the singular parts and $\gh$ additional moduli corresponding to the  addition of an arbitrary holomorphic differential to $v$.
Altogether, we get
\be
\dim \Hcal_{\gh}[d_1,\dots,d_\kk]= 4\gh-4 +\kk+\sum_{j=1}^\kk d_j
\la{dimabel}
\ee
The dimension of  $ \Hcal_{\gh}[d_1,\dots,d_\kk]$ coincides with the dimension of the relative homology group
\be
H_1(\Ch\setminus\{y_j\}_{j=1}^\kk,\;\{x_i\}_{i=1}^r)
\la{homgr}
\ee
A set of generators of this group can be chosen as follows:
\be
\{s_i\}_{i=1}^{\dim \Hcal_g[d_1,\dots,d_\kk]}= \left\{\{a_\a, b_\a\}_{\a=1}^{\gh}\;,\hskip0.3cm \{c_i\}_{i=2}^\kk, \;,\hskip0.3cm \{l_i\}_{i=1}^{r-1}\right\}
\la{defsi}
\ee
where $\{a_\a, b_\a\}$ form a Torelli marking on $\Ch$, $c_i$ are small counter-clockwise contours around $y_i$ and each contour $l_i$ connects 
$x_{r}$ with  $x_i$.

The  homology group dual to (\ref{homgr}) is 
\be
H_1(\Ch\setminus\{x_i\}_{j=1}^r,\;\{y_j\}_{j=1}^\kk)
\la{homgrdu}
\ee
and the set of generators dual to the set (\ref{defsi}) with the intersection index 
$$s_i^*\cdot s_j=\delta_{ij}$$ 
  is given by 
\be
s_i^*=\left\{\{-b_\a, a_\a\}_{\a=1}^g\;,\hskip0.3cm \{-\lt_i\}_{i=2}^\kk, \;,\hskip0.3cm \{\ct_i\}_{i=1}^{r-1}\right\}
\la{defdu}
\ee
where $\lt_i$ is the contour connecting the pole $y_1$ with $y_i$; $\ct_i$ is a small counter-clockwise contour around $x_i$.

The set of {\it homological}, or {\it period} coordinates on $\Hcal_{\gh}[d_1,\dots,d_\kk]$ is given by integrals of $v$ over the basis $\{s_i\}$ (\ref{defsi}):
\be
\Pcal_i=\int_{s_i} v\;,\hskip0.7cm i=1,\dots, \dim \Hcal_{\gh}[d_1,\dots,d_\kk]\;.
\ee

Introduce the following objects on $\Ch$: the prime-form $E(x,y)$, 
canonical bidifferential $B(x,y)$ 
(see for example \cite{Fay92}, Ch. II,  for the definition and properties of $E$ and $B$), holomorphic Abelian differentials $v_\a$ normalized via $\int_{a_\a}v_\b=\delta_{\a\b}$ and the
 period matrix $\Omega_{\a\b}=\int_{b_\a}v_\b$.
 
Choose a fundamental polygon  of $\Ch$  with vertex at $x_r$ and dissected along paths  connecting $x_r$ with poles $y_j$ (having only $x_r$ as common point);
denote the resulting simply connected domain by $\widetilde{\CC}$; on it we define  the "flat" coordinate 
\be
z(x)=\int_{x_r}^x v
 \la{flatco}\ee
 which can be used as local coordinate on $\Ch$ outside of zeros and poles of $v$.

 \begin{proposition}\cite{JDG,CMP}
 The following variational formul\ae\  for the period matrix $\Omega$ on the space $\Hcal_{\gh}[d_1,\dots,d_\kk]$ hold:
 \be
 \f{\p\Omega_{\a\b}}{\p \Pcal_j} =\int_{s_j^*} \f{v_\a v_\b}{v}\;.
 \la{varO}
 \ee
\end{proposition}

To present variational formul\ae\  for $v_\a$, $B$ and $E$ we need to  define their variations:  for 
 $v_\a$ we define
\be
\f{\p v_\a}{\p \Pcal_j} (x)= \f{\p}{\p\Pcal_j}\left(\f{v_\a(x)}{v(x)}\right)\Big|_{z(x)=const} v(x)\;.
\la{defvarv}
\ee
The result is a differential in $\widetilde {\CC}$ with  discontinuities across  all the dissecting cuts of $\Ch$ where the discontinuity is the addition of a constant depending which boundary component of the dissection we are crossing.
Analogously we define variations of $B(x,y)$ and $E(x,y)$ in $\Pcal_j$.

\begin{proposition}\cite{JDG,CMP}
The following variational formul\ae\  on the space $\Hcal_{\gh}[d_1,\dots,d_\kk]$ hold
\be
\f{\p v_\a(x)}{\p \Pcal_i} =\f{1}{2\pi i}\int_{t\in s_i^*} \f{v_\a(t)B(x,t)}{v(t)}\;,
\la{varva}
\ee
\be
\f{\p B(x,y)}{\p \Pcal_i} =\f{1}{2\pi i}\int_{t\in s_i^*} \f{B(x,t)B(t,y)}{v(t)}\;,
\la{varvB}
\ee

\be
\f{\p }{\p \Pcal_i} \log \le(E(x,y)\sqrt{v(x)}\sqrt{v(y)}\ri)=-\f{1}{4\pi i}\int_{t\in s_i^*} \f{1}{v(t)}\left[d_t\log\f{E(x,t)}{E(y,t)}\right]^2\;.
\la{varvE}
\ee
\end{proposition}

In the next section we show to deduce  variational formul\ae\  on spaces of spectral covers by restriction of the above ones.

On the subspace of  $\Hcal^0_{\gh}[d_1,\dots,d_\kk]$ of $\Hcal_{\gh}[d_1,\dots,d_\kk]$ defined by the vanishing of the residues of $v$ we 
 define the Bergman tau-function via  the system of differential equations \cite{JDG,CMP}:

\be
\f{\p}{\p\Pcal_j}\log\tau_B(\Ch,v)=\int_{s_j^*} \f{B^v_{reg}(x,x)}{v(x)}
\la{deftau}
\ee
where 
\be
B^v_{reg}(x,x)=\left(B(x,y)-\f{v(x)v(y)}{(\int_x^y v)^2}\right)\Big|_{x=y}\;.
\la{Bregdef}\ee
We refer to \cite{JDG,CMP} for explicit formula for $\tau_B$ and to \cite{MRL,CMP} for its properties and applications.

\section{Variational formul\ae\  on spaces of generalized Hitchin's covers}

We first  discuss the variations of the period matrix $\Oh$ of $\Ch$ on the moduli space $\Mcal_H^n[{\bf k}]$ of spectral covers:  these formul\ae\  are obtained by  pullback of the variational formul\ae\  on the space
$\Hcal_{\gh}({\bf d})$ of Abelian differentials on Riemann surfaces of genus $\gh$
where the vector ${\bf d}$ is given by
\be
{\bf d}=(k_1,\dots,k_1,k_2,\dots,k_2,\dots\dots, k_n,\dots,k_n)
\ee
where each $k_i$ is repeated $n$ times.
Thus in the context of previous section we have $k=nm$, and  the set of poles $\{y_j\}$ coincides with the set $\{y_j^{(s)}\}$,
$j=1,\dots,m$, $s=1,\dots,n$.

Assume that the branch points of $\Ch$, i.e., zeros of  the discriminant $W$ of (\ref{spcov}),  are also simple. We have  $(v)=D_{br}+D_0$
where $D_{br}$ is the divisor of ramification points of $\Ch$. The projection of $D_{br}$ on $\CC$ coincides with the divisor of the discriminant $W$: $\pi(D_{br})=(W)$. The projection of $D_0$ on $\CC$ coincides with the divisor of the $n$-differential $Q_n$:
$\pi(D_0)=(Q_0)$. Then   ${\rm deg} D_0=n(2g-2)+n\sum_{j=1}^n k_j$ i.e. ${\rm deg} D_{br}+{\rm deg} D_0=m$ as expected.
Let us enumerate their points as follows:
$$D_{br}=\{x_i\}_{i=1}^{{\rm deg} D_{br}}\;,\hskip0.7cm D_0=\{x_i\}_{{\rm deg} D_{br}+1}^{m}\;.$$

\noindent We now consider first  the  case of variations of the period matrix.

The map of $\Mcal_H^n[{\bf k}]$ to $\Hcal_{\gh}(d_1,\dots,d_{mn})$ is defined by assigning to a point of $\Mcal_H^n[{\bf k}]$ the
pair $(\Ch,v)$; for  a generic point of $ \Mcal_H^n[{\bf k}]$ all zeros of $v$  are simple.

\begin{theorem}\la{thvarO}
The variations of the period matrix $\Omega$ with respect to the coordinates (\ref{coord}) on $\Mcal_H^n[{\bf k}]$ are given by:
\be
\f{\p \O_{\a\b}}{\p A_\g}=-2\pi i \sum_{x_i\in D_{br}} \f{v_\g}{d\log(v/d\xi)}(x_i)\res {x_i}\f{v_\a v_\b}{v}\;,
\la{Oh1}
\ee
\be
\f{\p \O_{\a\b}}{\p C_j^{(s),l}}=-2\pi i \sum_{x_i\in D_{br}} \f{w_j^{(s),l}}{d\log(v/d\xi)}(x_i) \res{x_i}\f{v_\a v_\b}{v}\;,
\la{Oh10}
\ee
\be
\f{\p \O_{\a\b}}{\p C_j^{(s),1}}=-2\pi i \sum_{x_i\in D_{br}} \f{u_j^{(s)}}{d\log(v/d\xi)}(x_i)\res{x_i}\f{v_\a v_\b}{v}
\la{Oh100}
\ee
where in these formul\ae\  $\xi$ denotes a local coordinate on $\CC$ near $x_i$\footnote{We did not carry in the notation the dependence on $i$ for brevity of notation.}; the right-hand side of (\ref{Oh2}) is independent of the choice of these coordinates near $x_i$.
\end{theorem}

The formula (\ref{Oh1}) can be written alternatively in the following more familiar form:
\be
\f{\p \O_{\a\b}}{\p A_\g}=-2\pi i \sum_{x_i\in D_{br}}  \res{x_i}\f{v_\a v_\b v_\g}{d\xi \,d(v/d\xi)}
\la{Oh2}
\ee
and analogous  versions of (\ref{Oh10}) and (\ref{Oh100}) where  $v_\gamma$ is replaced by $w_j^{(s),\ell}$ and 
$u_j^{(s)}$, respectively.


On the submanifold $\Mcal_H^n[{\bf k}]$ of  $\Hcal_{\gh}({\bf d})$  we use the set of independent coordinates given by (\ref{coord}) so that 
the period coordinates (\ref{defsi}) on $\Mcal_H^n[{\bf k}]$ become functions of (\ref{coord}) defined implicitly by the condition that the moduli of the base curve $\CC$ are constants.

For the proof of Theorem \ref{thvarO} we need the following Lemma.
\begin{lemma}
Denote by  $s_k$  a contour from the list (\ref{defsi}) which does not coincide with 
with a contour connecting $x_r$ with $x_i$ with $x_i\in D_{br}$ ( a branchpoint). The derivatives of the integrals of $v$ over the  basis (\ref{defsi}) with respect to the coordinates (\ref{coord}) are then given by 
\be
\f{\p(\int_{s_k} v)}{\p z_j}=\int_{s_k}\f{\p v}{\p z_j}
\la{noend}\ee
where $z_j$ is any coordinate from the list (\ref{coord}) and the periods of the right-hand side are given by standard formul\ae\  taking into account (\ref{vA}), (\ref{vC1}),  (\ref{vC2}).

If $x_i$ is a branch point then the derivatives have the following additional contributions:
\be
\f{\p(\int_{x_r}^{x_i} v)}{\p A_\a}=\int_{x_r}^{x_i} v_\a-\f{v_\a}{d\log(v/d\xi)}(x_i)\;,
\la{bpA}
\ee
\be
\f{\p(\int_{x_r}^{x_i} v)}{\p C_j^{(s),\ell}}=\int_{x_r}^{x_i}  w_j^{(s),\ell}  -\f{w_j^{(s),\ell}}{d\log(v/d\xi)}(x_i)\;,
\la{bpC1}
\ee
\be
\f{\p(\int_{x_r}^{x_i} v)}{\p C_j^{(s),1}}=\int_{x_r}^{x_i} u_j^{(s)}-\f{u_j^{(s)}}{d\log(v/d\xi)}(x_i)\;
\la{bpC2}
\ee
where the coordinate $\xi$ is assumed to be invariant under the deformation. The expressions (\ref{bpA})-(\ref{bpC2})
are independent of the choice of local coordinate $\xi$ on $\CC$.
\end{lemma}

{\it Proof.}
We start from  (\ref{noend}): if the contour $s$ is  closed (i.e. coincides with one of $a$- or $b$-cycles or a small contour around one of $y_k^{(s)}$) then
the differentiation commutes with integration. If $s$ connects $x_r$ with another zero $x_j$ which  is not a branch point of $\Ch$ then $s$ can be projected on $\CC$, and in a  local coordinate on $\CC$ the integrand vanishes at both endpoints. 
Therefore, the differentiation commutes with integration in this case, too.

The only case when the dependence of the endpoint on the differentiation variable gives a non-trivial contribution is the case
when $s$ connects $x_r$ with one of the branch points $x_i$ of $\Ch$. Below we prove (\ref{bpA}); the proof of (\ref{bpC1}) and (\ref{bpC2}) is almost identical.

Let $x_i\in\Ch$ be a ramification point of $\Ch$ and $\xi_i=\xi(\pi(x_i))\in\CC$ be the corresponding critical value
in some local coordinate $\xi$ on $\CC$ which remains fixed under deformation of $\Ch$;
let $\zeta=\xi-\xi_i$ be a coordinate on $\CC$ vanishing at $\pi(x_i)$ (the coordinate $\zeta$ deforms when $\Ch$ varies). A suitable local coordinate on $\Ch$ near
$x_i$ can then be chosen to be $\zetah(x)=\zeta^{1/2}$.

Then the differentiation with respect to $A_\a$  of the endpoint also gives a contribution to $\p z_k/\p A_\a$ and we get 
\be
\f{\p (\int_{x_{r}}^{x_i}v)}{\p A_\a}=\int_{x_{r}}^{x_i} v_\a+ \f{\p\xi_i}{\p A_\a}\f{v}{d\xi}(\xi_i)
\la{zkaj1}
\ee
for  $k=1,\dots,{\rm deg} D_{br}$.

To compute the derivative  ${\p\xi_i}/{\p A_\a}$  we follow \cite{Ber} and we write $v(\xi)$ near $\xi_i$ in the form
\be
v=(a+b\sqrt{\xi-\xi_i}+\dots)d\xi
\ee
(recall that $v$ has simple zero in the local parameter $\sqrt{\xi-\xi_i}$ and  $d \xi$ has  already a simple zero).
Thus
\be
b=\f{d(v/d\xi)}{d\hat{\zeta}_i}\Big|_{\xi=\xi_i}=\f{d(v/d\xi)}{d\sqrt{\xi-\xi_i}}\Big|_{\xi=\xi_i}
\ee
and
\be
v_\a=\f{\p v}{\p A_\a}=\le( a_{A_\a}-\f{{\xi_i}_{A_\a}}{2\sqrt{\xi-\xi_i}}b+\dots \ri)d\xi\;.
\ee
Therefore,
\be
-\f{v_\a}{d\sqrt{\xi-\xi_i}}\Big|_{\xi=\xi_i}=b\f{\p\xi_i}{\p A_\a}
\ee
and,
\be
\f{\p\xi_i}{\p A_\a}=-\f{v_\a/d\zetah_i}{(v/d\xi)_{\zetah_i}}(x_i)\;.
\ee
Now (\ref{zkaj1}) takes the form
\be
\f{\p \int_{x_r}^{x_i} v}{\p A_\a}=\int_{x_{r}}^{x_i} v_\a-\f{v_\a/d\zetah_i}{[\log(v/d\xi)]_{\zetah_i}}(x_i)
\la{zkaj2}
\ee
for  $i=1,\dots,{\rm deg} D_{br}$.
This proves the lemma.
\QED
 
\noindent  {\bf Proof of  Theorem \ref{thvarO}}. 
Let us prove (\ref{Oh1}); the proofs of (\ref{Oh10}) and  (\ref{Oh100}) are parallel.

On the space $\Mcal_H^n[{\bf k}]$ the periods $B_\gamma$ and $\int_{x_r}^{x_i}v$ become functions
of $\{A_\g\}_{\g=1}^{\gh}$. Therefore  one  can compute derivatives of the period matrix 
using the chain rule: 
\be
\f{\p \O_{\a\b}}{\p A_\g}=\f{\p \O_{\a\b}}{\p A_\g}\Big|_{B_\g,(\int_{x_r}^{x_i}v)=const}+\sum_{\delta=1}^{\gh}\f{\p \O_{\a\b}}{\p B_\delta}\f{\p B_\delta}{\p A_\g} +\sum_{i=1}^{r-1}
\f{\p \O_{\a\b}}{\p (\int_{x_r}^{x_i}v)}\f{\p  (\int_{x_r}^{x_i}v)}{\p A_\g}
\la{varOA}
\ee
(since $A_\a$ and the  residues of $v$ are independent coordinates we omit the term involving these derivatives).
Using  (\ref{vA}), (\ref{bpA}) together with variational formul\ae\  (\ref{varO})
\be
\f{\p \O_{\a\b}}{\p A_\g} =-\int_{b_\g}\f{v_\a v_\b}{v}\;,\hskip0.7cm
\f{\p \O_{\a\b}}{\p B_\g} =\int_{a_\g}\f{v_\a v_\b}{v}\;,\hskip0.7cm
\f{\p \O_{\a\b}}{\p (\int_{x_r}^{x_i}v)} =2\pi i\res{x_i}\f{v_\a v_\b}{v}
\ee
(where  $x_i$ runs through the set of all zeros of $v$) 
we rewrite (\ref{varOA})  as follows:
\begin{eqnarray}
\f{\p \O_{\a\b}}{\p A_\g}=&\& \sum_{\delta=1}^{\gh}\left[-\left(\int_{a_\delta}v_\g\right)\left( \int_{b_\delta}\f{v_\a v_\b}{v}\right)
+\left(\int_{b_\delta}v_\g\right)\left( \int_{a_\delta}\f{v_\a v_\b}{v}\right)\right] + \nn\\
&\& 
+2\pi i \sum_{i=1}^{r-1}  \left(\int_{x_r}^{x_i}v_\g\right) \left(  \res{x_i}\f{v_\a v_\b}{v} \right)
-2\pi i \sum_{i=1}^{{\rm deg} D_{br}}    \f{v_\g}{d\log(v/d\zeta)}(x_i)    \res{x_i}\f{v_\a v_\b}{v}
\la{chainA}
\end{eqnarray}
Due to the Riemann bilinear identity the sum of the first three terms  in (\ref{chainA}) vanishes. The remaining terms give (\ref{Oh1}).

The formulas (\ref{Oh10}) and (\ref{Oh100}) are obtained in a similar way by applying Riemann bilinear identities  to the pairs  $(w_j^{(s),\ell }, \f{v_\a v_\b}{v})$ 
and $(u_j^{(s)}, \f{v_\a v_\b}{v})$, respectively.

We give below the computation leading to (\ref{Oh100}); the proof of (\ref{Oh10}) requires only minimal modifications.
Taking into account (\ref{bpC2}) we get (recall that all $a$-periods of $u_j^{(s)}$ vanish)
\be
\f{d \O_{\a\b}}{d C_j^{(s),1}}=\f{\p \O_{\a\b}}{\p C_j^{(s),1}}+\sum_{\delta=1}^{\gh}\left[
\f{\p \O_{\a\b}}{\p A_\delta}
\f{\p A_\delta}{\p C_j^{(s),1}}+
\f{\p \O_{\a\b}}{\p B_\delta}
\f{\p B_\delta}{\p C_j^{(s),1}} \right]+\sum_{k=1}^{r-1}
\f{\p \O_{\a\b}}{\p (\int_{x_r}^{x_i}v)}\f{\p  (\int_{x_r}^{x_i}v)}{\p C_j^{(s),1}}\;.
\la{varOC}
\ee
We have ${\p A_\delta}/{\p C_j^{(s),1}}=0$ since all $a$-periods of $u_j^{(s)}$ vanish; according to (\ref{varO}),
$$\f{\p \O_{\a\b}}{\p C_j^{(s),1}}=-2\pi i\int_{y_1^{(1)}}^{y_j^{(s)}} \f{v_\a v_\b}{v}\;,$$
which gives  
\be
\f{d \O_{\a\b}}{d C_j^{(s),1}}=-2\pi i\int_{y_1^{(1)}}^{y_j^{(s)}} \f{v_\a v_\b}{v}+\sum_{\delta=1}^{\gh}\left[\left(\int_{b_\delta} u_j^{(s)}\right)\left( \int_{a_\delta}\f{v_\a v_\b}{v}\right)\right]
\ee
\be
+2\pi i \sum_{i=1}^{r-1}  \left(\res{x_i} \f{v_\a v_\b}{v}\right)\left(\int_{x_r}^{x_i}u_j^{(s)}\right) 
-2\pi i \sum_{i=1}^{{\rm deg} D_{br}}    \f{u_j^{(s)}}{d\log(v/d\zeta)}(x_i)  \res{x_i}\f{v_\a v_\b}{v}\;.
\ee

Again, the Riemann bilinear identities applied to the pair of differentials of third kind  $u_j^{(s)}$ and $\f{v_\a v_\b}{v}$ prove the vanishing of the sum of all terms except the last one, leading to (\ref{Oh100}) (we notice that these two differentials have different positions of poles).

\subsection{Variations of Abelian differentials} 

Here we are going to use variational formul\ae\  (\ref{varva})-(\ref{varvE}) on moduli spaces of Abelian differentials to derive the following analogs of
Theorem \ref{thvarO}.

\begin{theorem}\la{thvarva}
The variations of canonical differentials  $v_\a$ with respect to coordinates (\ref{coord}) on $\Mcal_H^n[{\bf k}]$ are expressed by the following formul\ae:
\be
\f{\p v_\a(x)}{\p A_\g}=-\sum_{x_i\in D_{br}} \f{v_\g}{d\log(v/d\xi)}(x_i)\; \res{t=x_i}\f{v_\a(t) B(t,x)}{v(t)}
\la{va1}
\ee
\be
\f{\p v_\a(x)}{\p C_j^{(s),\ell}}=- \sum_{x_i\in D_{br}} \f{w_j^{(s),\ell }}{d\log(v/d\xi)}(x_i)\; \res{t=x_i}\f{v_\a(t) B(t,x)}{v(t)}
\la{va10}
\ee
\be
\f{\p v_\a(x)}{\p C_j^{(s),1}}=-  \sum_{x_i\in D_{br}} \f{u_j^{(s)}}{d\log(v/d\xi)}(x_i)\; \res{t=x_i}\f{v_\a(t) B(t,x)}{v(t)}
\la{va100}
\ee
where $\xi$ is a local coordinate on $\CC$ near $x_i$ as in Theorem \ref{thvarO}. The right-hand side of (\ref{Oh2}) is independent of the choice of these coordinates near $x_r$.
\end{theorem}
{\it Proof.} Let us show how to derive  (\ref{va1}) from the variational formul\ae\  (\ref{varva}). In comparison with the variational formul\ae\  for $\Omega$ proven above it is essential to carefully consider the  dependence of $v_\a$ on the point of $\Ch$, since the latter is deforming. Moreover, the  variation of $v_\a$ with respect to  $\Pcal_j$ used in (\ref{varva})
is defined by (\ref{defvarv}) where the ``flat'' coordinate is kept fixed,  while in (\ref{va1}) the differentiation is performed according to the rule (\ref{defder}) where $\xi$ is a local parameter lifted to $\Ch$ from $\CC$ which is assumed to be independent of moduli coordinates on $\Mcal_H^n[{\bf k}]$.

Taking into account these differences,  one can compute the left-hand side of (\ref{va1}) as follows. Let $f_\a(x)=v_\a/v$: then the left-hand side of (\ref{va1}) is rewritten as
$$
\f{\p v_\a(x)}{\p A_\g}=\f{\p (v f_\a(x))}{\p A_\g}=\f{\p  f_\a(x)}{\p A_\g}\Big|_{\xi(x)} v(x)+f_\a(x) \f{\p  v(x)}{\p A_\g}\Big|_{\xi(x)}
$$
$$
=  \f{\p  f_\a(x)}{\p A_\g}\Big|_{z(x)} v(x)+\f{\p f_\a(x)}{\p z(x)}\f{\p z(x)}{\p A_\g}\Big|_{\xi(x)} v(x)         +  f_\a(x) v_\g(x)
$$
\be
=      \f{\p  f_\a(x)}{\p A_\g}\Big|_{z(x)}       +\Acal_{\g}(x) d\left(\f{v_\a}{v}\right) +\f{v_\a v_\g}{v}(x)
\la{calc1}
\ee
where $\Acal_\g(x)=\int_{x_r}^x v_\g$ is the component $\g\in \{1,\dots, \gh\}$  of the Abel map.

The computation of the first term in (\ref{calc1}) can then be performed in complete analogy to (\ref{chainA}) with the differential $\f{v_\a v_\b}{v}(t)$ replaced by the 
differential $\f{1}{2\pi i}\f{v_\a(t)B(x,t)}{v(t)} $. Applying the Riemann bilinear relations to the  differentials   $v_\g$ and $ \f{1}{2\pi i}\f{v_\a(t)B(x,t)}{v(t)} $
we obtain the sum of terms entering  the right-hand side of (\ref{va1}) minus  the residue of $\f{1}{2\pi i}\f{v_\a(t)B(x,t)}{v(t)}\int_{x_r}^t v_\g$ at $t=x$.
This residue is equal to the sum of the last two terms in (\ref{calc1}) with  opposite sign. This gives (\ref{va1}).
The 
proofs of the  formul\ae\  (\ref{va10}) and (\ref{va100}) are parallel.

\subsection{Variations of prime-form and canonical bidifferential}

Variational formul\ae\  for $E(x,y)$ and $B(x,y)$ can be proven in parallel to Th.\ref{thvarva}.

As in the case of normalized canonical differential, we define the derivative of $B(x,y)$ and $E(x,y)$ with respect to any coordinate $z_i$ on
$\Mcal_H^n[{\bf k}]$ as 
\be
\f{\p B(x,y)}{\p z_i}=\f{\p}{\p z_i}\left(\f{B(x,y)}{d\xi(x)d\xi(y)} \right)d\xi(x)d\xi(y)
\ee
\be
\f{\p  E(x,y)}{\p z_i}=\f{\p}{\p z_i}\left({E(x,y)}[{d\xi(x)}{d\zeta(y)}]^{1/2} \right)[d\xi(x)d\zeta(y)]^{-1/2}
\ee
where $\xi(x)$ and $\xi(y)$ are local coordinates lifted to $\Ch$ from moduli-independent local coordinates on $\CC$, and these coordinates
remain fixed under differentiation.

\begin{theorem}\la{thvarB}
The variations of the canonical bidifferential $B(x,y)$ with respect to the coordinates (\ref{coord}) on $\Mcal_H^n[{\bf k}]$ are given by:  
\be
\f{\p B(x,y)}{\p A_\g}=- \sum_{x_i\in D_{br}} \f{v_\g}{d\log(v/d\xi)}(x_i)\;\res{t=x_i}\f{B(x,t) B(t,y)}{v(t)}
\la{B1}
\ee
\be
\f{\p B(x,y)}{\p C_j^{(s),\ell}}=- \sum_{x_i\in D_{br}} \f{w_j^{(s),\ell }}{d\log(v/d\xi)}(x_i)\; \res{t=x_i}\f{B(x,t) B(t,y)}{v(t)}
\la{B10}
\ee
\be
\f{\p B(x,y)}{\p C_j^{(s),1}}=-\sum_{x_i\in D_{br}} \f{u_j^{(s)}}{d\log(v/d\xi)}(x_i)\; \res{t=x_i}\f{B(x,t) B(t,y)}{v(t)}
\la{B100}
\ee
where $\xi$ is a local coordinate on $\CC$ near $x_i$; the right-hand side of (\ref{Oh2}) is independent on the choice of these coordinates near $x_r$.
\end{theorem}

\begin{theorem}\la{thvarE}
The variations of the prime-form with respect to coordinates (\ref{coord}) on $\Mcal_H^n[{\bf k}]$  are given by:
\be
\f{\p \log E(x,y)}{\p A_\g}=-\f{1}{2} \sum_{x_i\in D_{br}} \f{v_\g}{d\log(v/d\xi)}(x_i)\res{t=x_i}\f{1}{v(t)}\left[d_t\log\f{E(x,t)}{E(y,t)}\right]^2
\la{E1}
\ee
\be
\f{\p \log E(x,y)}{\p C_j^{(s),\ell}}=-\f{1}{2} \sum_{x_i\in D_{br}} \f{w_j^{(s),\ell}}{d\log(v/d\xi)}(x_i) \res{t=x_i}\f{1}{v(t)}\left[d_t\log\f{E(x,t)}{E(y,t)}\right]^2
\la{E10}
\ee
\be
\f{\p \log E(x,y)}{\p C_j^{(s),1}}=-\f{1}{2} \sum_{x_i\in D_{br}} \f{u_j^{(s)}}{d\log(v/d\xi)}(x_i)\res{t=x_i}\left\{\f{1}{v(t)}\left[d_t\log\f{E(x,t)}{E(y,t)}\right]^2\right\}
\la{E100}
\ee
where $\xi$ is a local coordinate on $\CC$ near $x_i$; the right-hand side of (\ref{Oh2}) is independent of the choice of these coordinates near $x_i$.
\end{theorem}

\subsection{The Bergman tau-function on spaces of  spectral covers}

The Bergman tau function on the moduli spaces of Abelian differentials is a natural higher genus analog of Dedekind's eta-function 
\cite{Annalen,JDG,MRL}. One can define two natural tau  functions associated to the moduli space of spectral covers; in the case of holomorphic $v$ these tau functions were introduced in \cite{Faddeev} and used to study the Picard group of the moduli spaces
(in  \cite{Faddeev} we considered the tau functions on universal spaces of spectral covers i.e. we allowed the base curve $\CC$ 
to vary). 


Here we restrict ourselves to the case of holomorphic $v$, namely, to moduli space $\Mcal_H^n$ of spectral covers of the ordinary Hitchin systems. In this case the equations for the Bergman tau functions take a similar form  to the  variational formul\ae\  for  the canonical objects considered above.

Denote the moduli space of  ordinary Hitchin's spectral covers by $\Mcal_H^n$; in this case all coefficients $Q_k$ of the equation 
(\ref{spcov}) are holomorphic $k$-differentials,  the genus of the spectral cover is  $\gh=n^2(g-1)+1$, the number of branch points is $p=n(n-1)(2g-2)$ and 
the total number of zeros of $v$ is $r=2\gh-2=p+2n(g-1)$. 
The differential $v$ is holomorphic, and the local
 coordinates on $\Mcal_H^n$  are given by the $a$-periods $A_\g=\int_{a_\g}v$.

Considering $\Mcal_H^n$ as a subspace of the space of holomorphic Abelian differentials with simple zeros $\Hcal_{\gh}$
we define the Bergman tau function  on $\Mcal_H^n$ by restriction of the Bergman tau function (\ref{deftau}) from $\Hcal_{\gh}$.

The resulting equations for $\tau_B$ (this tau function is defined by the formula (4.3) of \cite{Faddeev}) as function of periods $A_\g$ can be derived from (\ref{deftau}) in  analogy to  (\ref{Oh1}): 
\begin{proposition}
The Bergman tau-function $\tau_B(\Ch,v)$ on the space $\Mcal_H^n$ 
satisfies the following system of equations
\be
 \f{\p \log \tau_B}{\p A_\g}=-2\pi i \sum_{x_i\in D_{br}} \f{v_\g}{d\log(v/d\xi)}(x_i)\res{x_i}\left(\f{B^v_{reg}}{v}\right)
 -\f{\pi i}{8}\sum_{i=1}^{r} \res{x=x_i}\left(\f{v_\g(x)}{\int_{x_i}^x v}\right)\;.
 \la{dertauA}
\ee
\end{proposition}
 {\it Proof.} In parallel to (\ref{chainA}) we have, applying the chain rule to the equations (\ref{deftau}) and using
(\ref{vA}), (\ref{bpA}):
$$
\f{\p \log\tau}{\p A_\g}=\sum_{\delta=1}^{\gh}\left[-\left(\int_{a_\delta}v_\g\right)\left( \int_{b_\delta}\f{B_{reg}}{v}\right)
+\left(\int_{b_\delta}v_\g\right)\left( \int_{a_\delta}\f{B_{reg}}{v}\right)\right]
$$
\be
+2\pi i \sum_{i=1}^{r-1}  \left(\int_{x_r}^{x_i}v_\g\right) \left( \res{x_i}\f{B_{reg}}{v} \right)
-2\pi i \sum_{i=1}^{{\rm deg} D_{br}}    \f{v_\g}{d\log(v/d\zeta)}(x_i)     \res{x_i}\f{B_{reg}}{v}
\la{chaintau}
\ee
Using Riemann bilinear relations  the first sum in (\ref{chaintau}) equals to the sum of the residues  as follows
\be
-2\pi i \sum_{i=1}^{r}  \res{x_i} \left(\f{B_{reg}}{v}\int_{x_r}^{x}v_\g\right) .
\ee
{The main difference with the proof of the variational formula (\ref{Oh1}) for the period matrix is that the poles of $\frac{B_{reg}}{v}$ are of  order $3$ (as we see below) which leads to extra terms while computing the residues.}
Let us now represent $B_{reg}$ via difference of two projective connections \cite{JDG}:
\be
B_{reg}(x,x)=\f{1}{6}(S_B-S_v)
\ee
where $S_B$ is the Bergman projective connection (this projective connection is holomorphic; it equals to the constant term in the asymptotics of $B(x,y)$ 
on the diagonal equals $(1/6)S_B$) and $S_v$ is the meromorphic projective connection given by the Schwarzian derivative
$$
S_v(\xi)=\left\{\int^x v, \xi\right\}=\left(\f{v'}{v}\right)'-\f{1}{2}\left(\f{v'}{v}\right)^2
$$
in any local coordinate $\xi$ on $\Ch$. In a neighbourhood of a zero $x_i$ of $v$ we choose $\xi$ such that 
$v=\xi d\xi$; then near $x_i$ we have
$$
\f{1}{6}\f{S_B-S_v}{v}=\left(\f{1}{4\xi^3} +\f{1}{6\xi} S_B(\xi)\right) d \xi
$$
and 
$$
 \res{x_i} \left(\f{B_{reg}}{v}\int_{x_r}^{x}v_\g\right)= \left( \int_{x_r}^{x_i}v_\g\right) \res{x_i} \f{B_{reg}}{v} 
+\f{1}{8}\left(\f{v_\g}{d\xi}\right)_\xi(x_i)
 $$
 $$
 =\left( \int_{x_r}^{x_i}v_\g\right) \res{x_i} \f{B_{reg}}{v} 
+  \f{1}{16}\res{x=x_i}\left(\f{v_\g(x)}{\int_{x_i}^x v}\right)
$$
and (\ref{chaintau}) equals to
$$
-2\pi i \sum_{i=1}^{{\rm deg} D_{br}}    \f{v_\g}{d\log(v/d\zeta)}(x_i)     \res{x_i}\f{B_{reg}}{v} 
-\f{\pi i}{8}\sum_{i=1}^{r} \res{x=x_i}\left(\f{v_\g(x)}{\int_{x_i}^x v}\right)
$$
which gives (\ref{dertauA}).
\QED

\section{Higher order derivatives on $\Hcal_{\gh}$ and $\Mcal_H^n$}
\subsection{Space $\Hcal_{\gh}$}
The higher order  derivatives  with respect to moduli on the space $\Hcal_{\gh}$ can be obtained by a simple iteration of first derivatives. 

Let us consider first the multiple derivatives of the Bergman tau function.
Using the coordinates $\Pcal_i=\int_{s_i} v$, where $s_i\in H_1(\Ch,\{x_i\}_{i=1}^r)$, and referring to   (\ref{deftau}) and (\ref{varvB}) we find: 
\be
\f{\p^2}{\p\Pcal_i\p\Pcal_j}  \log \tau_B=\f{1}{2\pi i}{\rm symm}_{i,j}\int_{s_i^*}\int_{s_j^*}\f{B^2(x,y)}{v(x)v(y)}
\ee
where the symmetrization is  $1/2$ of the sum of   the $(ij)$ and $(ji)$ terms. The symmetrization  is necessary if the contours $s_i^*$ and $s_j^*$ have 
non-zero intersection index (see formulas (3.5) and (3.6) of \cite{JDG} for details 
about the extra term associated to the intersection point if the symmetrization is not 
assumed).

Further differentiation using  (\ref{varvB}) gives
\be
\f{\p^3}{\p\Pcal_i\p\Pcal_j\p\Pcal_k}  \log \tau_B=\f{2}{(2\pi i)^2}{\rm symm}_{(i,j,k)}\int_{x\in s_i^*}\int_{y\in s_j^*}\int_{t\in s_k^*}\f{B(x,y)B(x,t)B(t,y)}{v(x)v(y)v(t)}
\ee
where the symmetrization is again understood as averaging over the  6  permutations of
$(i,j,k)$.
The $n$th derivatives of $\tau_B$ are given by
\be
\f{\p^{(n)}}{\p\Pcal_{i_1}\dots\p\Pcal_{i_n}}  \log \tau_B=\f{1}{(2\pi i)^{n-1}}
{\rm symm}_{(i_1,\dots,i_n)}\int_{s_{i_1}^*}\dots \int_{s_{i_n}^*}
\Qcal_n(z_{1},\dots,z_{n})
\la{multtau}
\ee
where the  completely symmetric multi-differential $\Qcal_n$ is given by
\be
\Qcal_n(z_1,\dots,z_n)=2\sum_{all \;\;\Gamma}\frac{\prod_{j=1}^{n} B(z_{k_j},z_{k_{j+1}})}{\prod_{j=1}^n v(z_j)}
\ee
The sum runs over all $(n-1)!/2$ permutations $\Gamma=(k_1,\dots,k_n)$ of  $z_1,\dots,z_n$ which form a cycle of length $n$ (two such permutations are considered equivalent if they are related by cyclic permutation i.e. we do not distinguish between $(1234)$ and $(2341)$); $k_{n+1}$ is identified with $k_1$.

The multi-differentials $\Qcal_n(z_1,\dots,z_n)$ satisfy the relations 
\be
\f{\p}{\p\Pcal_i} \Qcal_n(z_1,\dots,z_n)=\f{1}{2\pi i}\int_{t\in s_i^*} \Qcal_{n+1}(z_1,\dots,z_n,t)\;.
\la{varW}
\ee

Another natural hierarchy of multi-differentials (although no longer completely symmetric) which 
are given by combinations of $B(x,y)$ can be obtained by differentiation of $B(x,y)$ itself.
Namely,  using the variational formula (\ref{varvB}) on the space $\Hcal_{\gh}$ we get

\be
\f{\p^{n}}{\p\Pcal_{i_1}\dots\p\Pcal_{i_n}}  B(x,y)=\f{1}{(2\pi i)^n}
{\rm symm}_{(i_1,\dots,i_n)}
\int_{s_{i_1}^*}\dots \int_{s_{i_n}^*}
\Rcal_{n+2}(x,z_{1},\dots,z_{n},y)
\la{multB}
\ee
where the multi-differentials $\Rcal_n$ with $n$ arguments are given by 
\be
\Rcal_n(z_1,\dots,z_n)=\sum_{all \;\;\tilde{\Gamma}}\frac{\prod_{j=1}^{n-1} B(z_{k_j},z_{k_{j+1}})}{\prod_{j=2}^{n-1} v(z_j)}
\ee
where in all products entering this sum, the indices $k_1, k_n$ are given by  $k_1=1$ and $k_n=n$; the sum goes over all  $(n-2)!$ paths $\tilde{\Gamma}$  connecting $z_1$ with $z_n$ which go only once through every vertex representing the other arguments $z_2,\dots,z_{n-1}$. 

The multi-differentials $\Rcal_n$ are symmetric  is  under permutations of the intermediate arguments $x_2,\dots,x_{n-1}$,
but not fully symmetric, in contrast to $\Qcal_n$.

The families of multi-differentials $\Qcal_n$ and $\Rcal_n$ as well as their variational formul\ae\ resemble the structures arising
in the framework of topological recursion of \cite{EO} (the genus of the base curve $\CC$ equals zero in the constructions of \cite{EO}).
Moreover, both $\Qcal_n$'s and $\Rcal_n$'s have second order poles when any two arguments coincide (in addition to generically simple poles at the branch points),  while the  multi-differentials $W_n$ of \cite{EO} have poles only at the ramification points of the cover.

The formula (\ref{multB}) implies the following expression  for the multiple derivatives of the period matrix $\O$ of $\Ch$ on  $\Hcal_{\gh}$:
\be
\f{\p^{(n)}}{\p\Pcal_{i_1}\dots\p\Pcal_{i_n}}  \O_{\a\b}=\f{1}{(2\pi i)^{n-1}} {\rm symm}_{(i_1,\dots,i_n)}\int_{s_{i_1}^*}\dots \int_{s_{i_n}^*}
\Rcal_n^{\a\b}(z_{1},\dots,z_{n})
\la{multO}
\ee
where
$$
\Rcal_n^{\a\b}(z_1,\dots,z_n)=\int_{x\in b_\a}\int_{y\in b_\b}\Rcal_{n+2}(x,z_1,\dots,z_n,y)
$$
or
\be
\Rcal_n^{\a\b}(z_1,\dots,z_n)=v_\a(z_1)v_\b(z_n)
\sum_{all \;\;\tilde{\Gamma}}\frac{\prod_{j=1}^{n-1} B(z_{k_j},z_{k_{j+1}})}{\prod_{j=1}^{n} v(z_j)}
\la{Rnab}\ee
where, as before, in all products entering this sum $k_1=1$ and $k_n=n$; the sum goes over all  $(n-2)!$ paths $\tilde{\Gamma}$ connecting $x_1$ with $x_n$ which go only once through every vertex representing other arguments $x_2,\dots,x_{n-1}$.

\subsection{The space $\Mcal_H^n$}

On the spaces of spectral covers  $\Mcal_H^n$ the multi-differentials $\Qcal_n$  are related to $\Qcal_{n+1}$ by formul\ae\  which can be derived from (\ref{varW}) in parallel to the proof of (\ref{va1}):
\be
\f{\p }{\p A_\g}\Qcal_n(z_1,\dots,z_n)=-\sum_{x_i\in D_{br}} \f{v_\g}{d\log(v/d\xi)}(x_i)\; 
\res{t=x_i}\left\{\f{1}{v(t)}\Qcal_{n+1}(z_1,\dots,z_n,t)\right\}
\la{varW1}
\ee
Similarly, the multi-differentials $\Rcal_n$ and $\Rcal_{n+1}$ are related by
\be
\f{\p }{\p A_\g}\Rcal_n(z_1,\dots,z_n)=-\sum_{x_i\in D_{br}} \f{v_\g}{d\log(v/d\xi)}(x_i)\; 
\res{t=x_i}\left\{\f{1}{v(t)}\Rcal_{n+1}(z_1,\dots,z_{n-1},t,z_n)\right\}
\la{varRn}
\ee
Integrating (\ref{varW1}) over two $b$-cycles with respect to $z_1$ and $z_2$ we get similar formul\ae\  for $\Rcal_n^{\a\b}$ (\ref{Rnab}).

While higher derivatives of the period matrix, tau-function and canonical bidifferential  on the space $\Hcal_{\gh}$ are given by a simple formul\ae\  
(\ref{multtau}), (\ref{multB}) and (\ref{multO})
their restriction to the space $\Mcal_H^n$ is much less trivial. 
As an example of such computation we find below the second derivatives of the period matrix.

\subsubsection{Second derivatives of $\O_{\a\b}$.}

The period matrix on the space  $\O_{\a\b}$ is known to be given by second derivatives of a single function (the "prepotential") 
$$
F_0=\f{1}{2}\sum_{\g=1}^{\gh} A_\g B_\g\;:
$$
\be
\Omega_{\a\b}=\f{\p^2 F_0}{\p A_\a \p A_\b}\;.
\la{OF}
\ee
We recall the  proof of  (\ref{OF}): using the relation $\p B_\g/\p A_\a=\Omega_{\a\b}$ we get
$$
\f{\p F_0}{\p A_\a}= \f{1}{2}\le (B_\a+\sum_{\g=1}^{\gh} A_\g \O_{\a\g}\ri )
$$
and
$$
\f{\p^2 F_0}{\p A_\a \p A_\b}=\O_{\a\b}+\f{1}{2}\sum_{\g=1}^{\gh} A_\g\f{\p \O_{\a\g}}{\p A_\b}\;.
$$
The last sum in this formula equals zero: indeed the formula (\ref{Oh2}) for $\p\O_{\a\b}/\p A_\g$ implies that this tensor is invariant under permutations of the indices $\a,\b,\g$ and thus we have
\be
\sum_{\g=1}^{\gh} A_\g\f{\p \O_{\a\g}}{\p A_\b}= \left(\sum_{\g=1}^{\gh} A_\g\f{\p }{\p A_\g}\right)\O_{\a\b}.
\ee
The last expression vanishes because it is the action of the scaling operator  $\mathbb E = \sum_{\g=1}^{\gh} A_\g\f{\p }{\p A_\g}$ generating the map $v \mapsto \lambda v $ ($\l \in \C^\times$) and the period matrix is clearly invariant under such rescaling. 
\hskip0.3cm

Due to (\ref{OF}) all higher derivatives of $\O_{\a\b}$ in $A_\g$'s are also completely symmetric with respect to all indices.
It is convenient to use the following  notation:
$$
y=\frac{v}{d\xi}
$$
which is a function defined on  the union of small disks on $\Ch$ around ramification points $x_i$  depending  on the choice of local parameter $\xi$ on $\CC$ near each branch point $x_i$. Since $v$ has a simple zero at $x_i$, in a neighbourhood of $x_i$ $y$ is a holomorphic function of the corresponding local parameter $\hat{\xi}_i=\sqrt{\xi-\xi(x_i)}$.

To compute second derivatives of $\O_{\a\b}$
on $\Mcal_H^n$ one can differentiate the formula (\ref{Oh2}) 
\be
\f{\p \O_{\a\b}}{\p A_\g}=-2\pi i \sum_{x_i\in D_{br}}  \res{x_i}\f{v_\a v_\b v_\g}{d\xi \,dy}
\ee
with respect to the coordinate $A_\d$ using (\ref{vA}) and (\ref{va1}).
Then due to (\ref{va1}) we have 
$$
\f{\p(dy)}{\p A_\delta}=d\left(\f{v_\delta}{d\xi}\right)
$$
{which has second order pole at $x_i$ ($d \xi$ has a simple zero at the ramification point,  and, therefore, $v_\d/d \xi$ has a  simple pole at $x_i$).}
We have then 
\begin{eqnarray}
\f{ \p^2\O_{\a\b} }{\p A_\delta \p A_\gamma}=&\& 2\pi i \!\!\!\! \sum_{x_i,x_j\in Br}
\res{t=x_i}\res{\tt=x_j}\left\{B(t,\tt)\f{v_\d(\tt)v_\g(\tt) v_\a(t)v_\b(t) +(perm\;\;of\;\;(\a,\b,\g))}{(dy\,d\xi)(t)\,(dy\,d\xi)(\tt)}\right\}+
\nn\\
&\& +2\pi i \sum_{x_i\in D_{Br}} \res{x_i}\left\{\f{v_\a v_\b v_\g}{(dy)^2 d\xi}d\left(\f{v_\d}{d\xi}\right)\right\}\;.
\la{derO21}
\end{eqnarray}

It is natural to treat the terms corresponding to $i=j$ in this double sum separately. First, we compute the residue at the first order pole arising from zero of $d\xi$ at $x_i$.
Namely, we have $d\xi=2\hat{\xi}d\hat{\xi}$ near each $x_i$ and using the notation $v_\a(x_i)=v_\a/d\hat{\xi} (x_i)$ we have
$$
\res{\tt=x_i}B(t,\tt)\f{v_\d(\tt)v_\g(\tt)}{(dy\,d\xi)(\tt)}=\f{1}{2}\f{B(t,x_i)v_\d(x_i)v_\g(x_i)}{y'(x_i)}
$$
where, for any differential $u$ on $\Ch$, the notation $u(x_i)$ is used to denote $(u/d\hat{\xi})(x_i)$ and the  prime denotes the derivative with respect to  $\hat{\xi}_i$.

The resulting expression has a third order pole at $x_i$  (arising from the double pole of $B(t,x_i)$ and the simple  zero of $d\xi$):
$$
\res{t=x_i}\left\{\f{B(t,x_i)}{(dy d\xi)(t)} v_\a(t)v_\b(t)\right\}=\f{1}{2} B_{reg}^{d\hat{\xi}}(x_i)\f{v_\a v_\b}{y'}(x_i) +
\f{1}{4}\left(\f{v_\a v_\b}{y'}\right)''
$$
where
$$
B_{reg}^{d\hat{\xi}}(x_i)=\left(B(x,y)-\f{d\hat{\xi}(x)d\hat{\xi}(y)}{(\hat{\xi}(x)-\hat{\xi}(y))^2}\right)\Big|_{x=y=x_i}
$$
is equal to $1/6$ of the  Bergman projective connection computed at $x_i$ in the  coordinate $\hat{\xi}$.

To compute the last residue in (\ref{derO21}) we notice that the corresponding expression has a  pole of third order at $x_i$.
Starting from 
$$
v_\d=\le (v_\d(x_i)+v_\d' \xh_i+ \f{v_\d''}{2}\xh^2+\dots\ri )d\xh_i\;,\hskip0.7cm d\xi=2\xh_i d\xh_i
$$
and
$$
\f{1}{d\xi}d\left(\f{v_\d}{d\xi}\right)=-\f{v_\d}{4\xh_i^3}+\f{v_\d''}{8\xh}+\dots\;.
$$
the last term in (\ref{derO21}) can be computed as follows:
$$
 \res{x_i}\left\{\f{v_\a v_\b v_\g}{(dy)^2}\left(-\f{v_\d}{4\xh_i^3}+\f{v_\d''}{8\xh}+\dots\right)\right\}
= -\f{1}{8}\left( \f{v_\a v_\b v_\g}{(y')^2}   \right)''v_\d+\f{1}{8}\f{v_\a v_\b v_\g v_\d''}{(y')^2} \;.
$$
Now the formula (\ref{derO21}) can be written as

$$
\f{1}{2\pi i}\f{ \p^2\O_{\a\b} }{\p A_\delta \p A_\gamma}=\f{1}{4}\sum_{x_i\neq x_j\in Br}
\left\{B(x_i,x_j)\f{v_\d(x_i)v_\g(x_i) v_\a(x_j)v_\b(x_j) +(cycl\;\;of\;\;(\a,\b,\g))}{y'(x_i)\,y'(x_j)}\right\}
$$
$$
+\f{1}{4}\sum_{x_i\in Br}\f{ B_{reg}^{d\hat{\xi}}(v_\a v_\b v_\g v_\d +(cycl (\a,\b,\g))}{y'^2}(x_i)
+\f{1}{8}\sum_{x_i\in Br}\left(\f{v_\g v_\d}{ y' }\left(\f{v_\a v_\b}{y'}\right)''(x_i)+(cycl (\a,\b,\g))\right)
$$
$$
+\f{1}{8}\sum_{x_i\in Br} \left[-\left( \f{v_\a v_\b v_\g}{(y')^2}   \right)''v_\d+\f{v_\a v_\b v_\g v_\d''}{(y')^2} \right]\;.
$$
 A  straightforward computation by expanding the derivatives above, shows that the sum of the last two terms is equal to
$$
-\f{1}{8}\sum_{x_i\in Br}\f{y'''}{y'^3}v_\a v_\b v_\g v_\d +
\f{1}{8}\sum_{x_i\in Br}\f{1}{y'^2}(v''_\a v_\b v_\g v_\d+ (cycl\;\; (\a,\b,\g,\d))\;.
$$

Therefore we get the following proposition:
\begin{proposition}
$$
\f{1}{2\pi i}\f{ \p^2\O_{\a\b} }{\p A_\delta \p A_\gamma}=\f{1}{4}\sum_{x_i\neq x_j\in Br}
\left[ B(x_i,x_j)\f{v_\d(x_i)v_\g(x_i) v_\a(x_j)v_\b(x_j) +(cycl\;\;(\a,\b,\g))}{y'(x_i)\,y'(x_j)}\right]
$$
\be
+\f{1}{8}\sum_{x_i\in Br} \left[\left(\f{6B_{reg}^{d\hat{\xi}}}{y'^2}-\f{y'''}{y'^3}\right)v_\a v_\b v_\g v_\d(x_i)+
\f{1}{y'^2}(v_\a'' v_\b v_\g v_\d + cycl\;\;(\a,\b,\g,\d))\right]\;.
\la{doubO}
\ee
\end{proposition}

The formula (\ref{doubO}) coincides with the expression obtained in Theorem 7.5 of \cite{Baraglia} using the framework of topological recursion of \cite{EO} (notice that $6B_{reg}^{d\hat{\xi}}$ is nothing but the Bergman projective connection $S_B$ which enters the formula (7.4) of \cite{Baraglia}.
\vskip0.5cm

To conclude,  we have shown that the deformation calculus on spaces of Hitchin's spectral covers can be naturally induced 
from 
a much more transparent deformation theory on   moduli spaces of holomorphic or meromorphic Abelian differentials on Riemann surfaces.
In consideration of the  close relationship between deformations of spectral covers and the theory of topological recursion of
\cite{EO},  it is natural to expect that the topological recursion itself could be a manifestation of a much less involved structure
associated to moduli spaces of Abelian differentials.\\[10pt]

\noindent {\bf Acknowledgements.}  The authors thank Jacques Hurtubise  for numerous illuminating discussions.
The work of M.B. was supported in part by the Natural Sciences and Engineering Research Council of Canada (NSERC) grant RGPIN-2016-06660.
The work of D.K. was supported in part by the NSERC grant
RGPIN/3827-2015. This work started during the workshop "Tau Functions of Integrable Systems and Their Applications"
at BIRS, Banff, September 03-07, 2018. The authors thank BIRS for hospitality and excellent working conditions.
We thank anonymous referees for useful comments and suggestions.

\end{document}